\documentclass[11pt]{article}

\usepackage[english]{babel}
\usepackage[latin1]{inputenc}
\usepackage{amsmath, latexsym}
\usepackage{amssymb}

\usepackage[authoryear]{natbib}
\usepackage{color}
    \definecolor{myred}{rgb}{0.5,0,0}
    \definecolor{myblue}{rgb}{0,0,0.75}
    \definecolor{mygreen}{rgb}{0,0.5,0}

\newif\ifpdf
\ifx\pdfoutput\undefined
   \pdffalse
\else
   \pdfoutput=1
   \pdftrue
\fi

\ifpdf
   \usepackage{graphicx}
  \usepackage[pdftex,
               pdftitle={Validation of internal rating systems and PD estimates},
               pdfsubject={From the book The Analytics of Risk Model Validation},
               pdfkeywords={},
               pdfauthor={Dirk Tasche},
               bookmarks=false,
               breaklinks=true,
               colorlinks=true,
               citecolor=myred,
               linkcolor=myblue,
               urlcolor=mygreen]{hyperref}
\else
   \usepackage{hyperref}
   \usepackage[dvips]{graphicx}
   \usepackage{rotating}
\fi

\numberwithin{equation}{section}

\newlength{\captionwidth}
\begin{document}

\title{Validation of internal rating systems and PD estimates}

\author{%
Dirk Tasche\thanks{Deutsche Bundesbank, Postfach
10 06 02, 60006 Frankfurt am
Main, Germany\newline E-mail: dirk.tasche@gmx.net}\ %
\thanks{The opinions expressed in this paper are those of the
author and do not necessarily reflect views of Deutsche
Bundesbank.}%
}

\date{May 2006}
\maketitle

%\begin{abstract}
%\end{abstract}

%\renewcommand{\thefootnote}{\arabic{footnote}}
%\setcounter{footnote}{0}
\setcounter{figure}{0}

%%%%%%%%%%%%%%%
% New section %
%%%%%%%%%%%%%%%
\section{Introduction}

This chapter elaborates on the validation requirements for rating systems
and probabilities of default (PDs)
which were introduced with the New Capital Standards
\citep[commonly called ``Basel II'', cf.][]{BC04}. We start in Section
\ref{se:regulatory} with
some introductory remarks on the topics and approaches that will be discussed
later on. Then we have a view on the developments in banking
regulation that have enforced the interest of the public in validation techniques.
When doing so, we put the main emphasis on the issues with \emph{quantitative validation}.
The techniques discussed here could be used in order to meet the
quantitative regulatory requirements. However, their appropriateness will
depend on the specific conditions under which they are applied.

In order to have a common ground for the description of the different
techniques, we introduce in Section \ref{se:basic} a theoretical framework that
will be the basis for the further considerations. Intuitively, a good rating
system should show higher probabilities of default for the less creditworthy
rating grades. Therefore, in Section \ref{se:monotonicity}, we discuss
how this monotonicity property is reflected in the theoretical framework from Section \ref{se:basic}.

In Section \ref{se:power}, we study the meaning of \emph{discriminatory power}
    and some tools for measuring it in some detail. We will see that there are tools
    that might be more appropriate than others for the purpose of regulatory validation
    of discriminatory power.
    The topic in Section \ref{se:calibration} is \emph{calibration} of rating systems.
    We introduce some of the tests that can be used for checking correct calibration
    and discuss the properties of the different tests.
    We then conclude in Section \ref{se:conclusions} with some comments on the question which tools might be
    most appropriate for quantitative validation of rating systems and probabilities
    of default.

%%%%%%%%%%%%%%%
% New section %
%%%%%%%%%%%%%%%
\section{Regulatory background}
\label{se:regulatory}

There is a long tradition of rating agencies grading
firms that issue bonds.
This aims primarily at facilitating the decision making of investors.
Very roughly, the rating methodology applied by the agencies could be
decribed as expert
    judgment that is based on hard as well as on soft facts.

Credit institutions have another fifty years long tradition of scoring borrowers.
This way, the credit institutions want to support credit decisions, i.e.\
decisions to grant credit or not. With regard to scoring, the predominant
methodology applied by the credit institutions could roughly be
described as use of statistically based score variables.

In the past, rating and scoring were regarded as being rather different
concepts. This was partly caused by the fact that rating and scoring
respectively are usually applied to populations with rather different characteristics. 
Ratings are
most frequently used for pricing of bonds issued by larger corporates.
Scores variables are primarily used for retail credit granting.

But also the background of the developers of rating methodologies
and scoring methodologies respectively is usually quite different. Rating
systems are often developed by experienced practitioners, whereas the
development of score variables tends to be conferred on experts in statistics.

With the rising of modern credit risk management, a more unified view of rating
and scoring has become common. This is related to the fact that today both rating
and score systems are primarily used for determining PDs of borrowers. PDs are the
crucial determinants for pricing and granting credit as well as for allocating
regulatory and internal capital for credit risks.

In \citet{BC04}, the Basel Committee on Banking Supervision recommends to
take rating and scoring as the basis for determining risk-sensitive regulatory
capital requirements for credit risks (Basel II). Compared to the Basel I standard,
where capital requirements are uniformly at eight percent in particular for corporate
borrowers irrespective of their creditworthiness, this is a major progress.

Credit institutions that apply the Basel II standardized approach
can base the calculation of capital requirements on agency ratings
which are called \emph{external ratings} in the Basel II wording. However,
at least in continental Europe, external ratings are available only for a
minority of the corporate borrowers. As a consequence, in practice the capital
requirements according to the standardized will not differ much from the
requirements according to the Basel I regime.

Credit institutions that are allowed to apply the internal ratings based (IRB) approach
will have to derive PDs from ratings or scores they have determined themselves. Such ratings
or scores are called \emph{internal ratings}. The PDs then are the main determinants
of the regulatory capital requirements. Note that in the IRB approach capital requirements
depend not only on PD estimates but also on estimates of LGD (loss given default) and
EAD (exposure at default) parameters. Validation of LGD and EAD estimates is not a topic
in this chapter.

As mentioned earlier, there are different ways to develop internal rating systems. On the one hand, there
is the traditional approach to rating which is primarily based on expert knowledge.
The number of rating grades is fixed in advance and assignments of grades are
carried out according to qualitative descriptions of the grades in terms of
economic strength and creditworthiness.

On the other hand, another -- also more or less traditional approach -- 
is scoring which is primarily based
on statistical methods. The first result then is a score variable that takes on
values on a continuous
scale or in a discrete range with many possible outcomes. The Basel II IRB approach
requires that the score values are then mapped on a relatively small number of
rating grades (at least seven non-default grades), but leaves the exact number of grades in the
institution's discretion.

Combinations of rating systems that are based on statistical models and rating
systems that are based on expert knowledge are called \emph{hybrid} models. All
kinds of combinations appear in practice, with quite different combination approaches.
Driven partly by an IRB approach requirement, hybrid models even seem to be predominant.
Often they occur in the shape of a statistical model whose output can be overridden
by expert decisions.

Among the rules on validation in the Basel II framework, two are particularly relevant
for statistically based quantitative validation \citep[see][§ 500]{BC04}.
\begin{itemize}
    \item ``Banks must have a robust system in place to validate the accuracy and consistency
of rating systems, processes, and the \emph{estimation of all relevant risk components}.''
    \item ``A bank
must demonstrate to its supervisor that the internal validation process enables it to assess
the \emph{performance of internal rating} and risk estimation systems consistently and meaningfully.''
\end{itemize}
The Basel Committee
    on Banking Supervision has established the Accord Implementation Group (AIG)
    as a body where supervisors exchange minds on implementation
    questions and provide general principles for the implementation of the Basel II framework.
    In particular, the AIG has proposed general principles for
    validation. Most of these principles are related to the validation process as such, and
    only some are relevant for quantitative validation. In the following list
     of principles \citep[cf.][]{BC05}
    the ones relevant for quantitative validation are emphasized.
\renewcommand{\labelenumi}{(\roman{enumi})}
\begin{enumerate}%\setlength{\itemsep}{0.5ex}
\item Validation is fundamentally about assessing the \emph{predictive
ability of a bank's risk
 estimates} and the use of ratings in credit processes.
    \item The bank has primary responsibility for validation.
 \item Validation is an iterative process.
\item There is \emph{no single validation method}. \item  Validation
should encompass both \emph{quantitative} and qualitative elements. \item
Validation processes and outcomes should be subject to independent
review.
\end{enumerate}
\renewcommand{\labelenumi}{(\arabic{enumi})}
Hence, in particular, the Basel Committee emphasizes that validation
is not only a quantitative statistical issue, but also involves an
important qualitative process-oriented component. This qualitative
component of validation is commonly considered equally if not more
important than the quantitative component. This chapter, however,
deals with the quantitative component of validation only\footnote{%
For more information on qualitative validation see, e.g., \citet{CEBS05}
.}.

Principle (i) of the AIG introduces the term ``predictive ability''.
    This is not a common statistical notion. It is not
    a priori clear whether it is related to well-known technical terms like ``unbiasedness'', ``consistency''
    and so on.
    However, there seems to be a consensus in the financial industry that
    ``predictive ability'' should be understood in terms of \emph{discriminatory power}
and correctness of \emph{calibration} of rating systems. We follow
this path of interpretation for the rest of the chapter.

Commonly\footnote{%
In this respect, we follow \citet{BC05b}.
}, discriminatory power is considered to be related to the discrimination between ``good'' and
``bad'' borrowers.
Additionally, there is a connotation of discriminatory power with the correctness
of the ranking of the borrowers by the rating system. While the importance of discriminatory
power is obvious, however, examining the ranking 
seems to be of secondary
importance, as in the end the ranking should be according to the size of the PD estimates.
Therefore, correct ranking will be reached as soon as the calibration of the rating system
is correct.
This is a consequence of the fact that correct calibration is usually understood as having found the
``true''
PDs (probabilities of default) for the rating grades. Correctness of the calibration
of a rating system may be understood as implementation of the Basel Committee's
requirement to assess
the quality of the \emph{estimation of all relevant risk components}. Checking discriminatory power may be
interpreted as implementation of the Basel Committee's
requirement to validate the \emph{performance of internal rating}.

With regard to quantitative validation, the Basel Committee states in §~501 of
\citet{BC04} ``Banks must regularly \emph{compare realised default rates with estimated PDs for each
grade} and be able to demonstrate that the realised default rates are within the expected
range for that grade.'' Hence there is a need for the institutions to compare PD estimates and
realized default rates at the level of single rating grades. Such a procedure is commonly called
\emph{back-testing}.

In §~502 the committee declares
``Banks must \emph{also use other quantitative validation tools} and comparisons with
relevant external data sources''. Thus, institutions are required to think about further
validation methods besides back-testing at grade-level.

In §~504 of \citet{BC04} the Basel Committee requires that ``Banks must have
well-articulated internal standards for situations where \emph{deviations
in realised PDs, LGDs and EADs from expectations become significant} enough to call the
validity of the estimates into question. These standards must take account of \emph{business cycles
and similar systematic variability} in default experiences.'' As a consequence,
institutions have to decide whether perceived differences of estimates and realized
values are really significant. Additionally, the committee expects that validation methods
take account of systematic dependence in the data samples used for estimating
the risk parameters PD, LGD and EAD.

In the retail exposure class, institutions need not apply fully-fledged rating systems
in order to determine PDs for their borrowers. Instead, they may assign borrowers to pools
according to similar risk characteristics. Obviously, the requirements for quantitative
validation introduced so far have to be modified accordingly.

%%%%%%%%%%%%%%%
% New section %
%%%%%%%%%%%%%%%
\section{Statistical background}
\label{se:basic}

\paragraph{Conceptual considerations.} The goal with this section on the statistical background
    is to introduce the model which will serve as the unifying framework
    for most of the more technical considerations in the part on validation techniques.
    We begin with some conceptual considerations.

    We look at rating systems in a \emph{binary classification}
    framework. In particular, we will show that the binary
    classification concept is compatible with the idea of having
    more than two rating grades. For the purpose of this chapter
    binary classification is
    understood in the sense of discriminating between the populations
    of defaulters and non-defaulters respectively.

    For the purpose of this chapter we assume that the
    score or rating grade $S$ (based on regression or other methods) assigned to a borrower summarizes
    the information which is contained in a set of co-variates
    (e.g.\ accounting ratios).
    Rating or score variable design, development or implementation\footnote{%
    The process of design and implementation should be subject to qualitative
    validation.
    } is not the
    topic of this presentation. We want to judge with statistical
    methods whether rating or scoring systems are appropriate for
    discrimination between ``good'' and ``bad'' and are well calibrated.

    With regard to calibration, at the end of the section we
    briefly discuss how PDs can be derived from the
    distributions of the scores in the population of the
    defaulters and non-defaulters respectively.
\paragraph{Basic setting.} We assume that with every borrower two random variables
    are associated. There is a variable $S$ that may take on
    values across the whole spectrum of real numbers. And there is another
    variable $Z$ that takes on the values $D$ and
    $N$ only. The variable $S$ denotes a score on a continuous scale
    that the institution has assigned to the borrower. It thus reflects
    the institutions's assessment of the borrower's creditworthiness. We
    restrict our considerations to the case of continuous scores
    since this facilitates notation and reasoning. The case of
    scores or ratings with values in a discrete spectrum can be
    treated much in a similar way. The variable $Z$ shows the state the borrower will have
    at the end of a fixed time-period, say after one year. This
    state can be \emph{default}, $D$, or
    \emph{non-default}, $N$. Of course, the borrower's
    state in a year is not known today. Therefore, $Z$ is a \emph{latent}
    variable.

The institutions's intention with the score variable $S$ is to
    forecast the borrower's future state $Z$, by relying on the
    information on the borrower's creditworthiness that is
    summarized in $S$. In this sense, scoring and rating are
    related to binary classification. Technically speaking,
    scoring can be called binary classification with a one-dimensional co-variate.

\paragraph{Describing the joint distribution of $\mathbf{(S,Z)}$ with conditional densities.}
As we intend statistical inference on the connections
    between the score variable $S$ and the default state variable $Z$,
    we need some information about the joint statistical
    distribution of $S$ and $Z$. One way to describe this joint
    distribution is by specifying first the marginal distribution of $Z$
    and then the conditional distribution of $S$ given values of $Z$.

    Keep in mind that the population of borrowers we are
    considering here is composed by the sub-population of
    future defaulters, characterized by the value $D$ of the state
    variable $Z$, and the sub-population of borrowers remaining
    solvent in the future, characterized by the value $N$ of the
    state variable $Z$. Hence, borrowers with $Z=D$ belong to the defaulters population,
borrowers with $Z=N$ to the non-defaulters population.

\begin{subequations}
The marginal distribution of $Z$ is very simple as it suffices
    to specify $p$, the \emph{total probability of default} (also called \emph{unconditional
    PD} in the whole
    population. Hence $p$ is the probability that the state variable
    $Z$ takes on the value $D$. It also equals 1 minus the probability that $Z$
    takes on the value $N$.
\begin{equation}\label{eq:totalPD}
  p \ =\ \mathrm{P}[Z = D]\ =\ 1- \mathrm{P}[Z = N].
\end{equation}
    Note that the conditional probability
    of default given that the state variable takes on $D$ is just 1
    whereas the conditional PD given that the state is $N$ is just
    0.
\refstepcounter{figure}
\begin{figure}[ht]
%  \begin{center}
\centering
  \parbox{12.0cm}{Figure \thefigure:
  \emph{Illustrative example of score densities conditional
  on the borrower's status (default or non-default).}}
\label{fig:1}%\\[2ex]
\ifpdf
    \resizebox{\height}{8.0cm}{\includegraphics[width=12.8cm]{dichten.pdf}}
\else
\begin{turn}{270}
\resizebox{\height}{12.8cm}{\includegraphics[width=8.0cm]{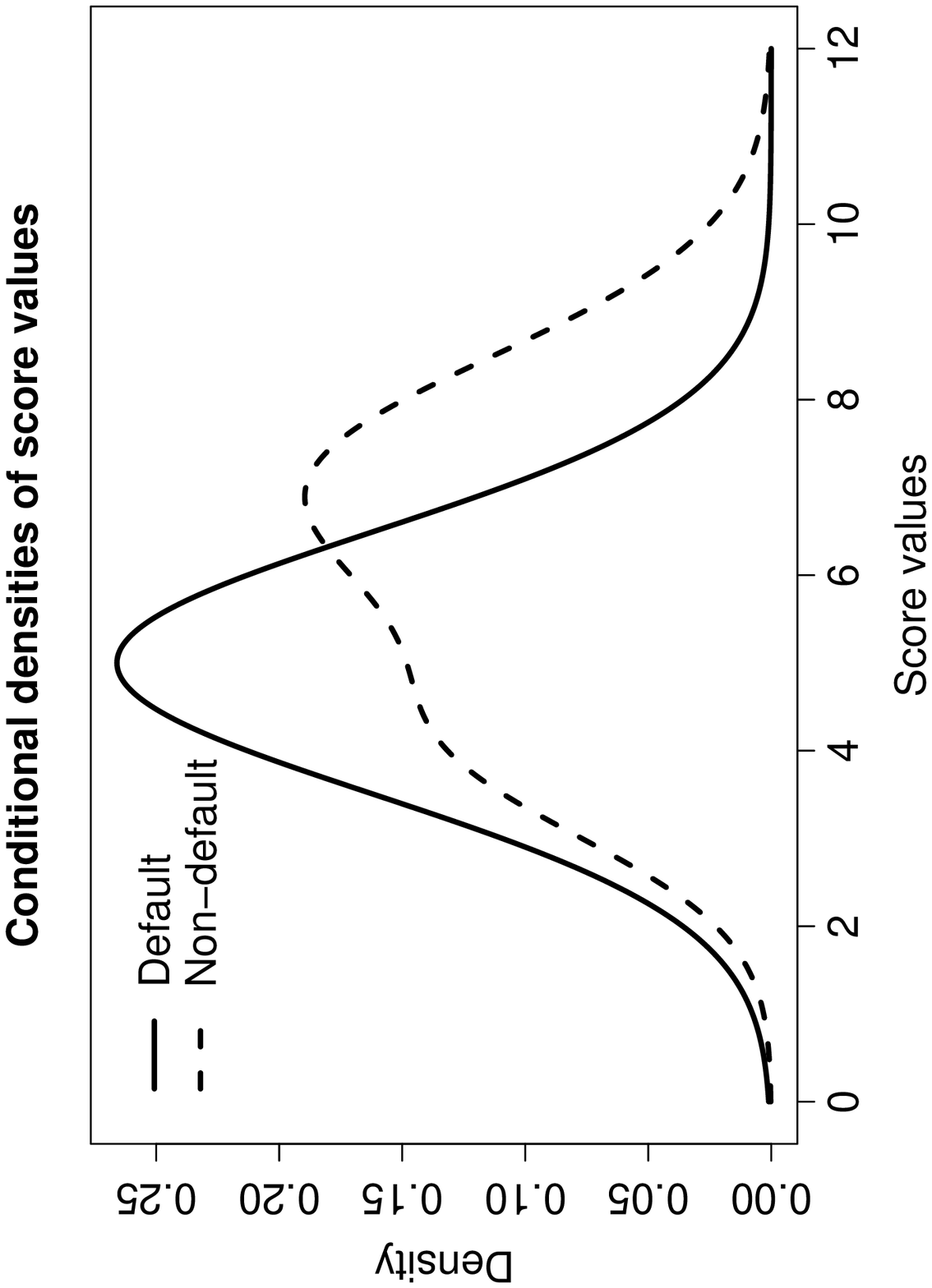}}
\end{turn}
\fi
%\end{center}
\end{figure}
As the score variable $S$ is continuous by assumption, its
    conditional distributions given the two values $Z$ can take on
    may be specified by \emph{conditional densities} $f_D$ and $f_N$
    respectively. Figure \ref{fig:1} illustrates how
    a plot of such conditional densities might look like. The probability that the score
    $S$ is not greater than some value $s$ given that
    the state variable $Z$ equals -- say -- $D$ can be
    expressed as an integral  of
    the density $f_D$.
\begin{equation}\label{eq:condDens}
    F_z(s)\ =\ \mathrm{P}[S\le s\,|\,Z=z]\ =\
  \int_{-\infty}^s f_z(u)\, d u, \quad z = D,N
\end{equation}
\end{subequations}

\paragraph{Describing the joint distribution of $\mathbf{(S,Z)}$
with conditional probabilities of default.}
Another, in a sense dual way of describing
    the joint distribution of $S$ and $Z$ is to specify for every value the
    score variable can take on the conditional probability $\mathrm{P}[Z=D\,|\,S =s]$ that
    the state variable $Z$ equals $D$. This is nothing but the
    conditional PD given the score. See Figure \ref{fig:2} for an example of
    how the graph of such a conditional PD function could look like.
    \begin{subequations}
\begin{equation}\label{eq:condPD}
\begin{split}
  s\, \mapsto \,\mathrm{P}[Z=D\,|\,S =s] & = \mathrm{P}[D\,|\,S =s] \\
   &=1-\mathrm{P}[N\,|\,S =s].
\end{split}
\end{equation}
In order to fully specify the
    joint distribution, in a second step then the unconditional
    distribution of the score variable $S$ has to be fixed, e.g.\ via
    an unconditional density $f$.
\begin{equation}\label{eq:uncondDens}
  \mathrm{P}[S\le s]\ =\
  \int_{-\infty}^s f(u)\, d u.
\end{equation}
\end{subequations}
Note the difference between the unconditional density $f$ of the score
variable $S$ on the one hand and
the on the state variable $Z$ conditioned densities $f_D$ and $f_N$ on the other hand.
The unconditional density $f$ gives the distribution of the scores in the whole population,
whereas $f_D$ and $f_N$ describe the score distribution on sub-populations only.
If the score variable $S$ really bears information about the default state, then the
three densities will indeed be different. If not, the densities might be
similar or even identical.
\refstepcounter{figure}
\begin{figure}[ht]
%  \begin{center}
\centering
  \parbox{12.0cm}{Figure \thefigure:
  \emph{Illustrative example of PD conditional on score values. Calculated
  with the densities from Figure \ref{fig:1} according to \eqref{eq:Bayes}. Total
  PD 10 percent.}}
\label{fig:2}%\\[2ex]
\ifpdf
    \resizebox{\height}{8.0cm}{\includegraphics[width=12.8cm]{BedPD.pdf}}
\else
\begin{turn}{270}
\resizebox{\height}{12.8cm}{\includegraphics[width=8.0cm]{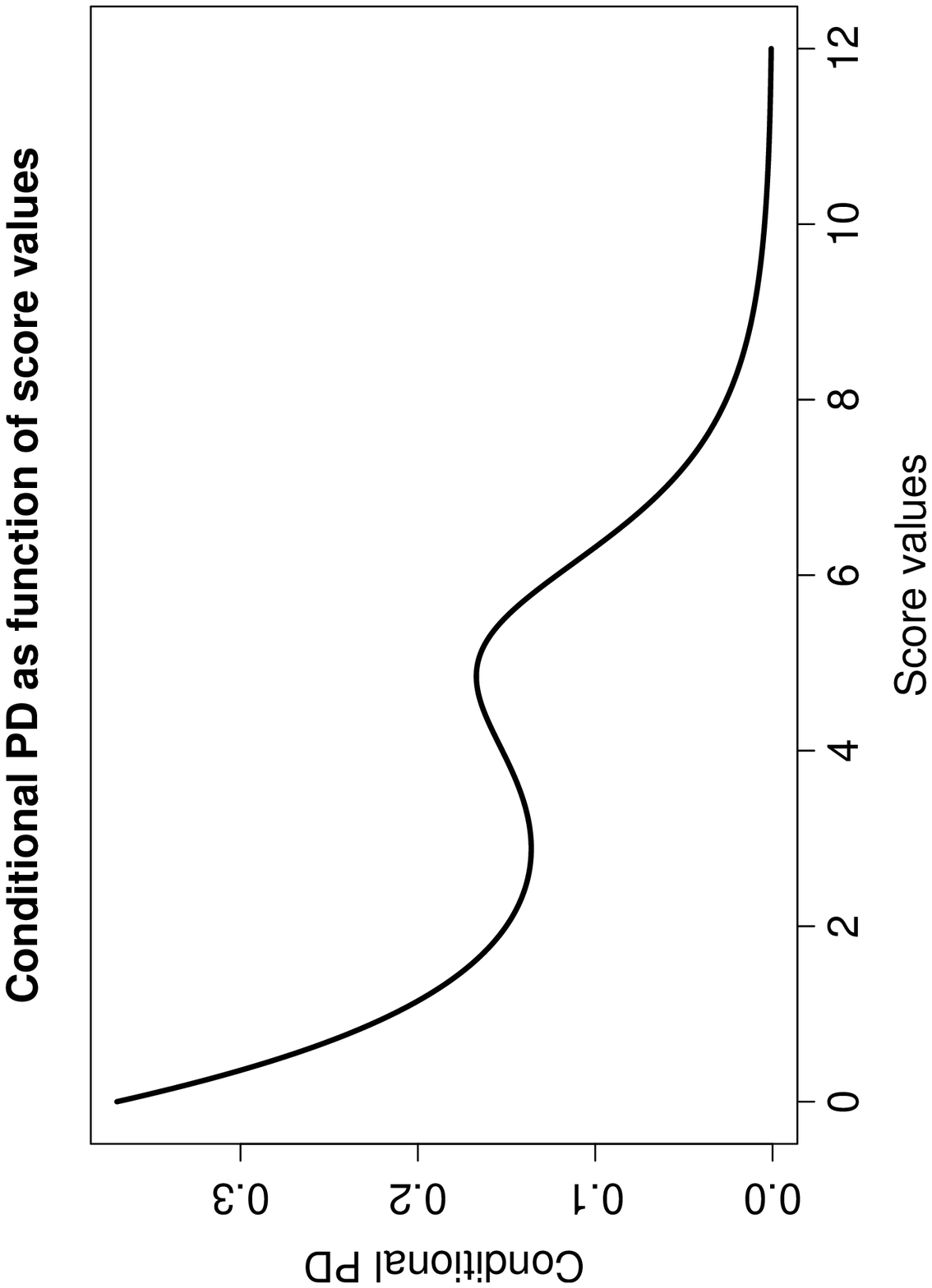}}
\end{turn}
\fi
%\end{center}
\end{figure}
By means of the densities $f$,  $f_D$ and $f_N$ the distributions
of all, the ``defaulting'' and the ``non-defaulting'' respectively borrowers'
score-variables are determined.

\paragraph{Equivalence of the both descriptions.} So far, we have seen two descriptions of the
    joint distribution of the score variable and the state variable
    which are quite different at first glance. However, thanks to Bayes' formula
    both descriptions are actually equivalent.

Suppose first that a description of the joint distribution
    of score and state by the total probability of default $p$ and
    the two conditional densities $f_D$ and $f_N$ according to
    \eqref{eq:totalPD} and \eqref{eq:condDens} is known. Then
    the unconditional score density $f$ can be expressed as
    \begin{subequations}
\begin{align}\label{eq:decomp_uncond_dens}
  f(s) & = p\,f_D(s) + (1-p)\,f_N(s), \\
\intertext{and the conditional
    PD given that the score variable takes on the value $s$ can
    be written as}
  \mathrm{P}[D\,|\,S=s] & = \frac{p\,f_D(s)}{f(s)}.\label{eq:Bayes}
\end{align}
\end{subequations}
Assume now that the unconditional score density $f$ in the sense
of \eqref{eq:uncondDens} and
    the function representing the conditional PDs given the scores
    in the sense of \eqref{eq:condPD} are known.
    Then the total PD $p$ can be calculated as an integral of
    the unconditional density $f$ and the conditional PD as
    \begin{subequations}
\begin{equation}\label{eq:uncondPD}
p \ =\ \int_{-\infty}^\infty \mathrm{P}[D\,|\,S=s]\,f(s)\,d s,
\end{equation}
and the both conditional densities of the score variable can be obtained via
\begin{equation}
\begin{split}
  f_D(s) & = \mathrm{P}[D\,|\,S=s]\,f(s) / p\quad\text{and}\\[1ex]
  f_N(s) & = \mathrm{P}[N\,|\,S=s]\,f(s) / (1-p).
\end{split}  
\end{equation}
    \end{subequations}
\paragraph{A comment on conditional PDs.} It is useful to keep
in mind that \eqref{eq:Bayes} is only one of several ways to
calculate conditional PDs. By definition, the conditional PD $\mathrm{P}[D\,|\,S]$
  can also be described as the best forecast of the default/non-default state variable
  $Z$ by a function of $S$ in the least squares sense, i.e.
  \begin{equation}\label{eq:least_squares}
    \mathrm{P}[D\,|\,S] \ =\ \arg \min_{Y = f(S),\ f\ \text{function}}
    \mathrm{E}\bigl[(Z - Y)^2\bigr].
\end{equation}
This means that
  the conditional PD can be regarded as the solution of an
  optimization problem where the objective is to approximate as best as possible 
  the state variable by some function of the score variable. Or,
  alternatively, it can be stated that given the information by the score $S$, there is
no better approximation (in the least squares sense) of the state
variable $Z$ than $\mathrm{P}[D\,|\,S]$. Intuitively this is quite clear, because obviously
a conditional PD of -- say -- 90 percent would indicate that
the borrower under consideration is close to default.

Note that due to the continuous distribution of $S$
for any $s \in \mathbb{R}$ we have $\mathrm{P}[S=s]=0$. Hence
$\mathrm{P}[D\,|\,S]$ is a conditional probability in the non-elementary
sense and has to be dealt with carefully.

\paragraph{Dealing with cyclical effects.} Recall from Section \ref{se:regulatory}
    that, for modelling, estimation and validation, institutions have
    to take account of ``business cycles
and similar systematic variability''.
     We consider
    here how such cyclical effects, expressed as additional dependence
    on time, can be incorporated in the model framework we have
    introduced.

    A first way to incorporate time-dependence is to
    assume that the conditional PDs $\mathrm{P}[D\,|\,S=\cdot]$
    provided by the model are
    constant over time and to admit time-dependence only via
    the unconditional score density $f$, i.e. $f$ is replaced
    by $f(\cdot, t)$.
    This corresponds to the
    so-called \emph{through-the-cycle} (TTC) rating philosophy, where
    rating grades are assumed to express the same degree of
    creditworthiness at any time and economic downturns
    are only reflected by a shift of the score distribution
    towards the worse scores.

    A second possibility to take account of time-dependence
    could be to assume that the conditional PDs are varying with time, i.e.
\begin{equation}
    \mathrm{P}[D\,|\,S=\cdot]\ =\ \mathrm{P}_t[D\,|\,S=\cdot].
\end{equation}
    This can be modelled with an assumption of constant conditional
    score densities $f_D$ and $f_N$, having only the total PD $p=p_t$ time-dependent.
    Via Bayes' formula \eqref{eq:Bayes}, the conditional PDs would then depend upon time too.
    This approach corresponds to the \emph{point-in-time} (PIT) rating philosophy
    according to which one and the same rating grade can reflect different
    degrees of creditworthiness, depending on the state of the economy.

\paragraph{The situation in practice.} If we think in terms of a statistically based
    rating or scoring system, we can expect that from the development of the score variable
    conditional densities of the scores in the two populations of
    defaulters and non-defaulters respectively are known. In some institutions,
    it is then practice to predict the total probability
    of default for the following period of time by means of a regression on macro-economic
    data. In such a situation, Bayes' formula is useful for deriving
    the conditional PDs given the values of the score variable. Note that
    we are then in the point-in-time context we have described above.
    In some cases, for instance when logit or probit regression is applied, the
    score variable itself can be interpreted as conditional PD. This would
    again be a case of a point-in-time rating philosophy.

    The popular software by Moody's-KMV provides a further example
    of this type. There one can also say that the score variable
    is identical with the conditional probability of default. The traditional
    Moody's (or S\&P or Fitch) ratings are commonly considered to be
    examples of the through-the cycle rating philosophy.

\paragraph{Mapping score values on rating grades.}
Due to our assumptions
and hence relevant for many score variables in practice,
the theoretical probability that the score variable $S$ takes on
some fixed score value $s$ is 0. As a consequence of this fact
PDs conditional on single scores $s$ cannot directly be back-tested
since there will not be many or even no observations of borrowers with score $s$.
In order to facilitate validation,
therefore, the Basel Committee on Banking Supervision decided
to admit for the IRB approach only rating systems with a finite number of
grades.
Thus, if a rating system is based on a continuous score variable,
a reasonable mapping of the scores on the grades must be constructed.
In the remaining part of this section, we show how such a mapping can be
constructed while taking into account the
intended PDs of the grades.

The main issue with mapping score values on rating grades
is to fix the criterion according to which the mapping
is defined. We consider
two different quantitative criteria.
The first criterion for the mapping we consider is
the requirement to have \emph{constant PDs over time}.

To describe the details of the corresponding
mapping exercise, assume that $k$ non-default rating grades
have to be defined. Grade 1 denotes the grade that
indicates the highest creditworthiness. Increasing
PDs $q_1 < \ldots< q_{k-1}$ have been fixed. Assume additionally,
that the conditional PD $\mathrm{P}[D\,|\,S=s]$ as a function of the score values is
decreasing in its argument $s$. We will come back to a justification of
this assumption
in Section \ref{se:monotonicity}.

Given the model framework introduced above,
the theoretical conditional PD given that the
score variable takes on a value equal to or higher than a fixed
limit $s_1$ can be determined. This observation also holds
for the PD conditional on the event that the score variable
takes on a value between two fixed limits. Speaking technically, it is possible,
by proceeding recursively,
to find limits $s_1 > s_2 > \ldots > s_{k-1}$ such that
\begin{subequations}
\begin{equation}\label{eq:limit1}
\begin{split}
  q_1 = \mathrm{P}[D\,|\,S \ge s_1] & = \frac{p \int_{s_1}^\infty f_D(u)\,du}
  {\int_{s_1}^\infty f(u)\,du} \quad \text{and}\\[1ex]
  q_i = \mathrm{P}[D\,|\,s_{i-1} > S \ge s_i] & = \frac{p \int_{s_i}^{s_{i-1}}
  f_D(u)\,du}
  {\int_{s_i}^{s_{i-1}} f(u)\,du} \quad \text{for}\ i = 2, \ldots, k-1. 
\end{split}  
\end{equation}
If a borrower has been assigned a score value $s$ equal to
or higher than the limit $s_1$, he or she receives the
best grade $R(s)=1$.
In general, if the score value is less than a limit $s_{i-1}$
but equal to or higher than the next limit $s_i$ then the intermediate grade
$R(s)=i$ is assigned. Finally, if the borrower's score value is less than
the lowest limit $s_{k-1}$ then he or she receives the worst grade $R(s)=k$.
\begin{equation}\label{eq:mapping_1}
  R(s) \ =\
  \begin{cases}
    1 & \text{if }s \ge s_1, \\
    i & \text{if }s_{i-1} > s \ge s_i,\ i = 2, \ldots, k-1,\\
    k & \text{if }s_{k-1} > s.
  \end{cases}
\end{equation}
The PD of the worst grade $k$ cannot be fixed in advance. It turns
out that its value is completely determined by the PDs of the better grades.
This value, the conditional PD given that the rating grade
is $k$, can be calculated in a way similar to the calculations of the conditional
PDs for the better grades in \eqref{eq:limit1}. As the
result we obtain
\begin{equation}\label{eq:PD_k}
  \mathrm{P}[D\,|\,R(S) = k] \ =\ \mathrm{P}[D\,|\, s_{k-1} > S]\ =\
p \,\frac{\int\limits_{-\infty}^{s_{k-1}}
  f_D(u)\,du}
  {\int\limits_{-\infty}^{s_{k-1}} f(u)\,du}.
\end{equation}
\end{subequations}
As $s \mapsto \mathrm{P}[D\,|\,S=s]$ decreases,
$r \mapsto \mathrm{P}[D\,|\,R(S)=r]$ is an increasing function, as
should be expected intuitively.

The second criterion for the mapping we consider is
the requirement to have a \emph{constant distribution of
the good borrowers across the grades over time}. The intention
with such a criterion might be to avoid major
shifts with respect to ratings in the portfolio when
the economy undergoes a downturn.

Assume hence that a number $k$ of non-default grades (grade 1 best) and shares
 $0 < r_1, \ldots, r_k$,
  $\sum_{i=1}^k r_i=1,$ of good borrowers in the grades have been fixed. Assume
  again additionally that high score
  values indicate high creditworthiness. It turns out that it is then possible,
  again by proceeding recursively, to
  find limits
 $s_1 > s_2 > \ldots > s_{k-1}$ such that
\begin{subequations} 
\begin{equation}\label{eq:limit_pop}
\begin{array}{rcccl}
  r_1 & =& \mathrm{P}[S \ge s_1\,|\,N] &=& \int\limits_{s_1}^\infty f_N(s)\,d s,\\[2ex]
  r_i & =& \mathrm{P}[s_{i-1} > S \ge s_i\,|\,N] &=& \int\limits_{s_i}^{s_{i-1}}
  f_N(s)\,d s,\
i = 2,\ldots, k-1.
\end{array}
\end{equation}
The mapping of the scores on grades in this case is again defined by \eqref{eq:mapping_1}.
Since $\sum_{i=1}^k r_i=1$ this definition implies immediately
\begin{equation}\label{eq:grade_k}
  \mathrm{P}[R(S) = k\,|\,N] \ =\ \mathrm{P}[S < s_{k-1}\,|\,N] \ = \
  r_k.
\end{equation}
\end{subequations}
Note that in both cases of mapping criteria we have considered,
in order to keep constant over time the PDs and the shares
of good borrowers respectively, the limits $s_1\ldots, s_{k-1}$ must be periodically updated.

%%%%%%%%%%%%%%%
% New section %
%%%%%%%%%%%%%%%
\section{Monotonicity of conditional PDs}
\label{se:monotonicity}

The mapping procedures described in the last part of Section \ref{se:basic}
work under the assumption that the conditional PD given the score
$\mathrm{P}[D\,|\,S=s]$ is a function that decreases in its argument
$s$. As Figure \ref{fig:2} demonstrates this need not be the case.
Are there any reasonable conditions such that monotonicity of the conditional
PDs given the score is guaranteed?

An answer\footnote{%
This section is based on \citet{Tasche02}.
} to this question, in particular, will provide
us with a justification of the monotonicity assumption which underlies
Equations \eqref{eq:limit1} to \eqref{eq:PD_k}. This assumption
is needed to ensure that the proposed mapping procedure
for having constant PDs over time really works.

We discuss the question in the context of a \emph{hypothetical decision problem}.
Assume that we consider a borrower chosen at random and have been informed
about his or her score value. But, of course, we do not yet know whether the
borrower will default or not. How could we infer the value of his or her default
state variable $Z$?
In formal terms: suppose that a realization $(s,z)$ of $(S,Z)$ has been sampled.
$s$ is observed, $z$ is not yet visible.
Is $z = N$ or  $z = D$?

One way to come to a decision would be to fix some set $A$ of score values
such that we infer default, $D$, as state if the borrower's score value is in $A$.
If the value
of the score were not in $A$, we would conclude that the state is $N$ for non-default, i.e.
\begin{equation}\label{eq:decision}
\begin{split}
  s \in A &\ \Rightarrow \ \text{Conclusion }z = D \\
  s \notin A &\ \Rightarrow \ \text{Conclusion }z = N.
\end{split}
\end{equation}
How should we choose the \emph{acceptance set} $A$? A convenient way to deal with the problem
of how to find $A$ is to have recourse to statistical test theory.

Then the problem can be stated as having to
discriminate between the conditional score distributions $\mathrm{P}[S\in \cdot\,|\,D]$ on
the defaulters and $\mathrm{P}[S\in \cdot\,|\,N]$ on the
non-defaulters sub-populations respectively.
Thus we have to decide whether the borrower under consideration
stems from the defaulters population or from the non-defaulters
population.
The key concept for solving this problem is to have as
objective a high certainty in the case of a decision to reject
the presumption that the borrower is a future defaulter.

Formally, we can state this concept as follows:
the null hypothesis is that the borrower is a future
defaulter, or, equivalently, that his or her state
variable takes on the value $D$.
The alternative hypothesis is that the borrower's state
variable has got the value $N$.
\begin{itemize}
    \item \emph{Null hypothesis:} $\mathrm{P}[S\in \cdot\,|\,D]$, i.e.\ $z=D$.
    \item \emph{Alternative:} $\mathrm{P}[S\in \cdot\,|\,N]$, i.e.\ $z=N$.
\end{itemize}
We conduct a statistical test on the null hypothesis
``state equals $D$'' against the alternative ``state equals $N$''.
Hence, our decision could be wrong in two ways.
The so-called type I error would be to reject ``state
equals $D$'' although the state is actually $D$.
The so-called type II error would be to accept
``state equals $D$'' although ``state equals $N$'' is true.
\begin{itemize}
    \item \emph{Type I error}: erroneously rejecting $z=D$.
    \item \emph{Type II error}: erroneously accepting $z=D$.
\end{itemize}
In order to arrive at an optimal decision criterion,
the probabilities of the two possible erroneous decisions
have to be considered:
The probability of the type I error is the probability
under the defaulters' score distribution that a borrower's
score will \emph{not} be an element of the acceptance set $A$.
\begin{subequations}
\begin{equation}
    \label{eq:typeI} \mathrm{P}[\text{Type I error}] \ =\ \mathrm{P}[S\notin A\,|\,D].
\end{equation}
In contrast, the probability of the type II error is the probability
under the non-defaulters' score distribution that a borrower's
score will be an element of the acceptance set $A$.
\begin{equation}
    \label{eq:typeII} \mathrm{P}[\text{Type II error}] \ =\ \mathrm{P}[S\in A\,|\,N].
\end{equation}
\end{subequations}
The type I error probability
is usually limited from above by a small constant.
Common values are 1 or 5 percent, but
we will see in Section \ref{se:power} that for the purpose of validation also higher
values make sense. Having bounded the type I error probability from above, the
objective is to minimize the type II error probability.
\begin{equation}\label{eq:error_opt}
\begin{split}
  \mathrm{P}[\text{Type I error}] & \le\ \text{(small) constant} \\
  \mathrm{P}[\text{Type II error}] & \ \text{as small as possible}.
\end{split}
\end{equation}
Note that $1-\mathrm{P}[\text{Type II error}] = \mathrm{P}[S\notin A\,|\,N]$
is called the \emph{power of
the test} we are conducting.

The optimal solution for the decision
problem \eqref{eq:error_opt} is provided by the well-known Neyman-Pearson lemma.
We state here briefly a slightly simplified version. See
\citet{CasellaBerger01} or other textbooks
on statistics for a more detailed version of the lemma.
Let now $\alpha$ denote a fixed bound for the type I error
probability, say $\alpha$ equal to 5 percent. Such an $\alpha$ is
called \emph{confidence level}.

The first step in stating the \emph{Neyman-Pearson lemma} is to introduce another
random variable, the so-called \emph{likelihood ratio}. It is obtained
by applying the ratio of the non-defaulters and the defaulters
conditional score densities $f_N$ and $f_D$ respectively as a function
to the score variable itself. The second
step is to determine the $1-\alpha$-quantile $r_\alpha$ of
the likelihood ratio.
\begin{subequations}
\begin{equation}\label{eq:quantile}
    r_\alpha \ =\ \min\bigl\{r \ge 0:\,\mathrm{P}\bigl[\frac{f_N}{f_D}(S) \le
r\,\big|\,D \bigr]\ge 1-\alpha\bigr\}.
\end{equation}
In the case of any reasonable
assumption on the nature of the conditional continuous score distributions
this can be done by equating $1-\alpha$ and the probability that
the likelihood ratio is not higher than the quantile, i.e.
\begin{equation}
    1-\alpha \ =\
\mathrm{P}\bigl[\frac{f_N}{f_D}(S) \le r_\alpha\,\big|\,D \bigr],
\end{equation}
\end{subequations}
and then solving
the equation for the quantile $r_\alpha$. Having found the quantile of the likelihood ratio,
the decision rule
``Reject the hypothesis that the future state is $D$ if the likelihood ratio
is greater than the quantile'' is optimal among all the decision rules
that guarantee a type I error probability not greater than $\alpha$. Hence, formally
stated, the decision rule
\begin{subequations}
  \begin{align}\label{eq:decision_rule}
  \frac{f_N}{f_D}(S)&\ >\ r_\alpha \quad \iff \quad \text{rejecting }D\\
\intertext{minimizes the type II error under the condition}
\mathrm{P}[\text{Type I error}] &\ \le\ \alpha.\label{eq:condition}
\end{align}
\end{subequations}
As a consequence, for any decision rule of the shape
\begin{subequations}
\begin{equation}
  S \notin A \quad \iff \quad \text{rejecting }D
\end{equation}
and with $\mathrm{P}[S \notin A\,|\,D]\le \alpha$ we have
\begin{equation}\label{eq:gleich}
  \mathrm{P}\bigl[\frac{f_N}{f_D}(S) \le r_\alpha\,\big|\,N
\bigr] \ \le \ \mathrm{P}[S \in A\,|\,N].
\end{equation}
\end{subequations}
In other words,  any optimal test of $D$ (``defaulter'') against
$N$ (``non-defaulter'') at level
$\alpha$ looks like the likelihood ratio test, i.e.\ from
(\ref{eq:gleich}) follows 
\begin{equation}\label{eq:optimal}
A \ =\ \bigl\{s: \frac{f_N(s)}{f_D(s)} \le
r_\alpha\bigr\}.
\end{equation}
Actually, \eqref{eq:optimal} is not only the optimal decision
criterion for discriminating between defaulters and non-defaulters,
but also provides an answer to
the original question of when the conditional PD given the scores
is a monotonous function.

\paragraph{Cut-off decision rules.} To explain this relation we need a further definition.
 A score variable $S$ is called to be of \emph{cut-off type} with respect to
  the distributions
  $\mathrm{P}[S\in\cdot\,|\,D]$ and $\mathrm{P}[S\in\cdot\,|\,N]$, if for
  every type I error probability $\alpha$ a decision rule of half-line shape
  \begin{subequations}
\begin{align}\label{eq:cut_I}
  S > r_\alpha & \quad \iff \quad \text{rejecting }D\\
\intertext{or for every $\alpha$ a rule of half-line shape}
  S < r_\alpha & \quad \iff \quad \text{rejecting }D \label{eq:cut_II}
\end{align}
\end{subequations}
is \emph{optimal} in the sense of minimizing the type II error probability
under the constraint \eqref{eq:condition}. Decision rules as in \eqref{eq:cut_I} and
\eqref{eq:cut_II} are called
\emph{cut-off rules}.

By \eqref{eq:optimal}, we can now conclude that
\begin{center}\emph{%
 the score variable
  $S$ is of cut-off type with respect to $\mathrm{P}[S\in\cdot\,|\,D]$ and
  $\mathrm{P}[S\in\cdot\,|\,N]$, if and only if the likelihood ratio
  $s\mapsto f_N(s) / f_D(s)$ is monotonous.}
  \end{center}
Note that, for any score variable, its corresponding likelihood ratio is of
cut-off type.

\paragraph{Conclusions for practical applications.}
  Bayes' formula \eqref{eq:Bayes} shows that the likelihood
  ratio is monotonous if and only if the conditional PD $s \mapsto \mathrm{P}[D\,|\,S=s]$
  is monotonous. There are some theoretical examples
  where the likelihood ration is indeed monotonous: for instance  when
  both conditional densities $f_N$ and $f_D$ are normal densities, with
  equal standard deviation.

  Unfortunately, in practice, monotonicity of the likelihood ratio or the conditional PD is hard to
  verify.
  However, from economic considerations can be clear that cut-off decision rules
  for detecting potential defaulters are optimal. This may justify the assumption of
  monotonicity. If, however, non-monotonicity of the likelihood ratio is visible from graphs
    as in Figure \ref{fig:2}, the reliability of the
    score variable may be questioned. This yields a first example for a validation criterion for
score variables, namely is the likelihood ratio monotonous or not?

%%%%%%%%%%%%%%%
% New section %
%%%%%%%%%%%%%%%
\section{Discriminatory power of rating systems}
\label{se:power}

The following section is devoted to studying the question of
    how discriminatory power can be measured and tested.
    We have seen in Section \ref{se:basic} that the statistical
    properties of a score variable can to a high extent be expressed by the conditional
    densities of the score variable on the two populations of the defaulters and non-defaulters
    respectively. Another, closely related way of characterization is by means
    of the conditional probability of default given the score values.
    With this observation in mind, discriminatory power can roughly be described
    in technical terms as discrepancy of the conditional densities, as variation
    of the conditional PD or as having the conditional PDs as close as possible to
    100 percent or 0 percent.

    Many statistical tools are available for measuring discriminatory power
    in one of these ways. We will consider a selection of tools that enjoy
    some popularity in the industry.
    A first major differentiation among the tools can be applied according to
    whether or not their use involves estimation of the total (or portfolio-wide or unconditional)
    probability of default. If this is necessary, the tool can be applied only to
    samples with the right proportion of defaulters.
    If estimation of the total PD is not involved, the tool can also be
    applied to non-representative samples. This may be important in particular
    when the power shall be estimated on the development sample of a rating system.
    The presentation of tools in the remaining part of this section closely follows the
    presentation in Chapter III of \citet{BC05b}. Of course, the presented list of tools for measuring discriminatory
    power is not exhaustive. Which tool should be preferred may strongly depend
    on the intended application. As a consequence, various scientific disciplines like statistics
    in medicine, signal theory, or weather forecasting in the course of time suggested
    quite different approaches on how to measure the discriminatory power
    of classification systems.

\paragraph{Cumulative Accuracy Profile (CAP).} The Cumulative Accuracy Profile (or CAP)
is a useful graphical tool for investigating the discriminatory power of rating systems.
Recall from \eqref{eq:condDens} the notions $F_N$ and $F_D$ for the distribution functions
of the score variable on the non-defaulters' and defaulters' respectively populations.
Denoting by $p$ as in Section \ref{se:basic} the total (portfolio-wide) probability of default,
it follows from \eqref{eq:decomp_uncond_dens} that the unconditional distribution function
$F(s)$ of the score variable can be written as
%
%\begin{subequations}
\begin{equation}\label{eq:score_uncond_dist}
    F(s)\ =\ \mathrm{P}[S \le s] = (1-p)\,F_N(s) + p\,F_D(s).
\end{equation}
The equation of the \emph{CAP function} is then given by
\begin{equation}\label{eq:CAP}
  CAP(u)\ =\ F_D\bigl(F^{-1}(u)\bigr), \quad u \in (0,1).
\end{equation}
%\end{subequations}
%
The graph of the \emph{l} can either be drawn by plotting all the
points $(u, CAP(u))$, $u\in (0,1)$ or by plotting all the points
$(F(s), F_D(s))$, $s \in \mathbb{R}$. The latter parametrization of
the CAP curve can still be used when the score distribution function $F$ is
not invertible. Figure \ref{fig:3} shows examples of how a CAP curve
may look like. The solid curve belongs to the score variable whose
two conditional densities
are shown in Figure \ref{fig:1}. A curve like this could occur in practice.
The two other curves in Figure \ref{fig:3} correspond to the so-called
\emph{random} (dotted line) and \emph{perfect} (dashed curve) respectively score variables.
In case of a random score variable the two conditional densities $f_D$ and $f_N$
are identical. Such a score variable has no discriminatory power at all. In case
of a perfect score variable the densities $f_D$ and $f_N$ have disjoint supports, i.e.
\begin{equation}\label{eq:support}
     \{s:\ f_D(s) > 0\} \cap \{s:\ f_N(s) > 0\}\ =\ \emptyset.
\end{equation}
\eqref{eq:support} implies that the realizable range of the scores of the defaulting borrowers
and the realizable range of the scores of the non-defaulting borrowers are disjoint, too. As a
consequence, perfect discrimination of defaulters and non-defaulters would be possible.
\refstepcounter{figure}
\begin{figure}[ht]
%  \begin{center}
\centering
  \parbox{12.0cm}{Figure \thefigure:
  \emph{Illustrative example of CAP curves of life-like, random and
  perfect score variables as explained in main text. For the score
  variable based on the conditional densities
  shown in Figure \ref{fig:1}. Total PD 10 percent. $AR = 0.336$.}}
\label{fig:3}%\\[2ex]
\ifpdf
    \resizebox{\height}{10.0cm}{\includegraphics[width=10.0cm]{CAP.pdf}}
\else
\begin{turn}{270}
\resizebox{\height}{10.0cm}{\includegraphics[width=10.0cm]{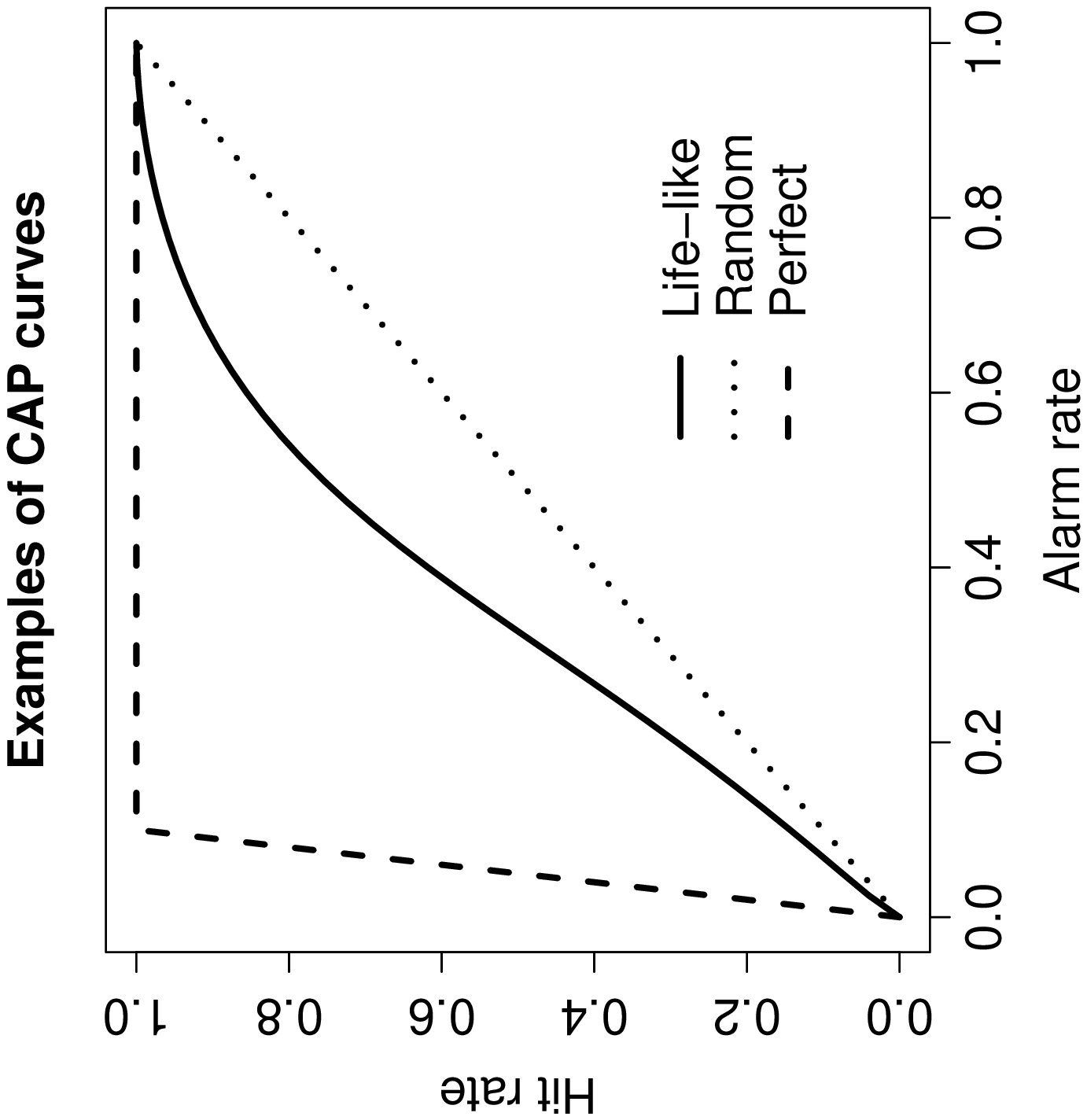}}
\end{turn}
\fi
%\end{center}
\end{figure}

In the context of CAP curves, $F(s)$ is called \emph{alarm rate} associated
with the score level $s$ 
and $F_D$ is called \emph{hit rate} associated
with the score level $s$. These names indicate what happens if
all the borrowers with a score equal to or less
than some fixed threshold $s$ are considered suspect of default (cut-off rule in the
sense of \eqref{eq:cut_I} and \eqref{eq:cut_II}). The hit rate then reflects
which portion of the defaulters will be detected by this procedure. The alarm rate gives the
portion of the whole population which will raise suspicion of being prone to default.
From these observations it follows that $100\,CAP(u)$\%
indicates the percentage of default-infected borrowers that
are found among the first (according to their scores) $100\,u$\% of all borrowers.
A further consequence is that the
``perfect'' curve in Figure \ref{fig:3} corresponds to a score variable which is of
cut-off type in the
sense of Section \ref{se:monotonicity}.

It seems unlikely that any rating system or score variable from practice will receive a
CAP-curve like that from the perfect rating system since this would indicate
that it will enable its owners to detect defaulters with certainty.
Similarly, it is not very likely to observe in practice a rating system with
zero power as a CAP-curve identical to the diagonal would indicate.
However, if a rating system is developed for a certain portfolio and
then is used on a completely different one, a very low discriminatory
power can be the result of such a procedure.

It is easy to show that the CAP function is related to the
conditional probability of default given the score via its
derivative scaled by the total probability of default.
\begin{equation}\label{eq:CAP_deriv}
  CAP'(u)\ =\ \mathrm{P}[D\,|\, S = F^{-1}(u)]\,/\,p.
\end{equation}
Recall from \eqref{eq:Bayes} that by Bayes' formula
the conditional probability of default given the score can be represented as
a ratio involving the conditional score densities. As a consequence of \eqref{eq:CAP_deriv}
and that representation,
the stronger is the growth of $CAP(u)$ for $u$ close to 0 (implying the conditional
PD being close to 1 for low scores) and the weaker
is the growth of $CAP(u)$ for $u$ close to 1 (implying the conditional PD
being close to 0 for high scores), the more differ the conditional
densities and the better is the discriminatory power of the underlying score
variable.

\paragraph{Accuracy Ratio (AR).} From Figure \ref{fig:3}, it is intuitively
clear that the area between the diagonal line and the CAP curve
can be considered a measure of discriminatory power. The random score variable
receives area 0, and the life-like score variable obtains an area greater than 0 but
less than the area of the perfect score variable. The area between the
diagonal and the CAP-curve of the life-like rating system (solid line) can be
calculated as the integral from 0 to 1 of the CAP function \eqref{eq:CAP} minus 1/2.
The area between the curve of the perfect score variable and the diagonal
is given by $1/2-p/2$ when $p$ denotes the total probability of default.

The so-called \emph{Accuracy Ratio} (AR) (also \emph{Gini-coefficient}) is defined
as the ratio of the area between the CAP-curve and the diagonal and the area
between the perfect CAP curve and the diagonal, i.e.
\begin{subequations}
\begin{equation}
  AR\ = \ \frac{2 \int_0^1 CAP(u)\, d u - 1}{1-p}.
\end{equation}
Alternatively, the Accuracy Ratio can be described as the difference
of two probabilities. Imagine that two borrowers are independently selected at random,
one from the defaulters population and the other from the non-defaulters population.
The first probability is the probability of the event to observe a higher
score for the non-defaulting borrower. The subtracted probability is the probability
of the event that the defaulting borrower has the higher score. Then
\begin{equation}
  AR\  =\ \mathrm{P}[S_D < S_N] -
\mathrm{P}[S_D > S_N],
\end{equation}
where $S_N$ and $S_D$ are independent and distributed
according to $F_N$ and $F_D$ respectively.
\end{subequations}
Obviously, if we assume that in general non-defaulters have the higher scores,
we will expect that the first probability is higher than the second as is
also indicated by the graph in Figure \ref{fig:3}.

From Figure \ref{fig:3} we can conclude that the discriminatory power
of a rating system will be the higher, the larger its Accuracy Ratio is. This
follows from \eqref{eq:CAP_deriv}, since a large Accuracy Ratio
implies that the PDs for the low scores are large whereas the PDs for the high
scores are small.

\paragraph{Receiver Operating Characteristic (ROC).}
The Receiver Operating Characteristic (or ROC) is another graphical tool for
investigating discriminatory power. Define, additionally to the notions of hit rate
and alarm rate from the context of the CAP curve, the \emph{false alarm rate}
associated with the score level $s$ as the conditional probability $\mathrm{P}[S \le s\,|\,N]
= F_N(s)$ that the score of a non-defaulting borrower is less than or equal to
this score level. Then the false
alarm rate reflects the
portion of the non-defaulters population which will be under wrong suspicion when
a cut-off rule with threshold $s$ is applied.
The equation of the \emph{ROC function} is now given by
\begin{subequations}
\begin{equation}\label{eq:ROC}
  ROC(u)\ =\ F_D\bigl(F_N^{-1}(u)\bigr), \quad u \in (0,1).
\end{equation}
The graph of the \emph{s} can either be drawn by plotting all the
points $(u, ROC(u))$, $u\in (0,1)$ or by plotting all the points
$(F_N(s), F_D(s))$, $s \in \mathbb{R}$. The latter parametrization of
the ROC curve can still be used when the conditional score distribution function $F_N$ is
not invertible. In contrast to the case with CAP curves, constructing
ROC curves does not involve
estimation of the total (or portfolio-wide) probability of default.
Figure \ref{fig:4} shows examples of how a ROC curve
may look like. The solid curve belongs to the score variable whose
two conditional densities are shown in Figure \ref{fig:1}. The dotted and
the dashed curves correspond to the random score variable and the perfect
score variable respectively as in the case of the CAP curves in Figure \ref{fig:3}.
\refstepcounter{figure}
\begin{figure}[ht]
%  \begin{center}
\centering
  \parbox{12.0cm}{Figure \thefigure:
  \emph{Illustrative example of ROC curves of life-like, random and
  perfect score variables as explained in main text. For the life-like score
  variable based on the conditional densities
  shown in Figure \ref{fig:1}. $AUC=0.668$.}}
\label{fig:4}%\\[2ex]
\ifpdf
    \resizebox{\height}{10.0cm}{\includegraphics[width=10.0cm]{ROC.pdf}}
\else
\begin{turn}{270}
\resizebox{\height}{10.0cm}{\includegraphics[width=10.0cm]{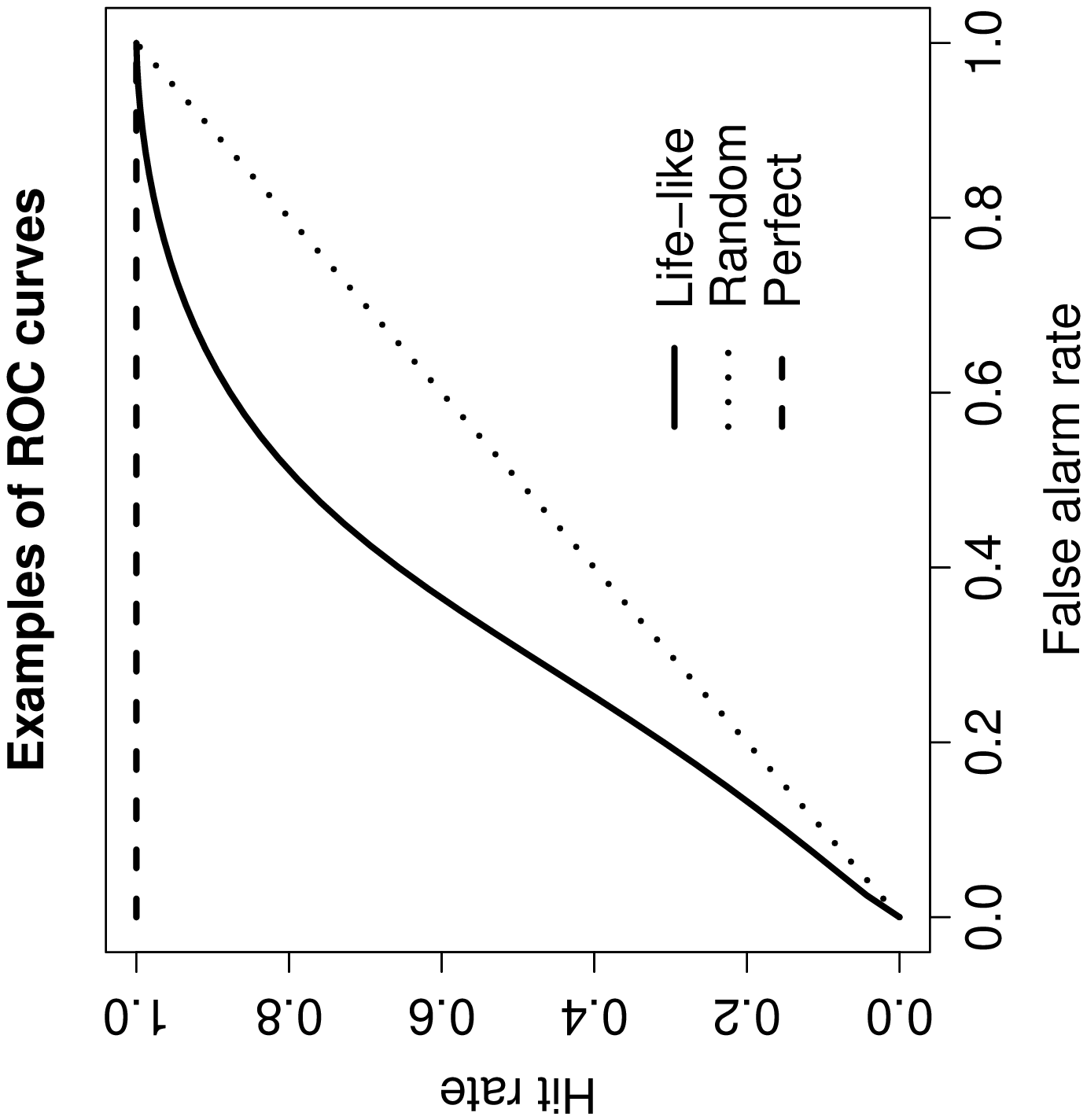}}
\end{turn}
\fi
%\end{center}
\end{figure}
As a result of this definition, $100\,ROC(u)$\% indicates the percentage of
default-infected borrowers that
have been assigned a score which is lower than the highest score of the first (according to
their scores)
$100\,u$\% non-defaulters. Alternatively the points on the ROC curve can be
characterized as all pairs
of type I error probability and power (see Section \ref{se:monotonicity})
that can arise when cut-off rules are applied
for testing the hypothesis ``non-default'' against the alternative ``default''.

The derivative of the ROC curve turns out to be closely related to the
likelihood ratio which was already mentioned in the context of the Neyman-Pearson
lemma in Section \ref{se:monotonicity}.
\begin{equation}\label{eq:ROC_deriv}
    ROC'(u)\ =\ \frac{f_D(F_N^{-1}(u))}{f_N(F_N^{-1}(u))}, \qquad u \in (0,1).
\end{equation}
Hence, the stronger is the growth of $ROC(u)$ for $u$ close to 0 and the weaker
is the growth of $ROC(u)$ for $u$ close to 1, the more differ the conditional
densities and the better is the discriminatory power of the underlying score
variable.
\end{subequations}

From Section \ref{se:monotonicity} we know that the score variable is of cut-off type and
hence optimal in a test-theoretic sense if and only if the likelihood ratio is
monotonous. Via \eqref{eq:ROC_deriv} this is also equivalent to the CAP curve
being concave or convex as concavity and convexity mean that the first derivative
is monotonous. If high scores indicate high creditworthiness, the conditional score
density $f_D$ is small for high scores and large for low scores and the conditional
score density $f_N$ is large for high scores and small for low scores. As a conclusion,
the ROC curve of an optimal score variable need to be concave in the case where
high scores indicate high creditworthiness. While the lack of concavity of the
solid curve in Figure \ref{fig:4} is not very clear, from the graph of its derivative
according to \eqref{eq:ROC_deriv} in Figure \ref{fig:5} the lack of monotonicity
is obvious.
\refstepcounter{figure}
\begin{figure}[ht]
%  \begin{center}
\centering
  \parbox{12.0cm}{Figure \thefigure:
  \emph{Derivative of the ROC curve given by the solid line in Figure \ref{fig:4}. For the life-like score
  variable based on the conditional densities
  shown in Figure \ref{fig:1}.}}
\label{fig:5}%\\[2ex]
\ifpdf
    \resizebox{\height}{8.0cm}{\includegraphics[width=12.8cm]{ROC_deriv.pdf}}
\else
\begin{turn}{270}
\resizebox{\height}{12.8cm}{\includegraphics[width=8.0cm]{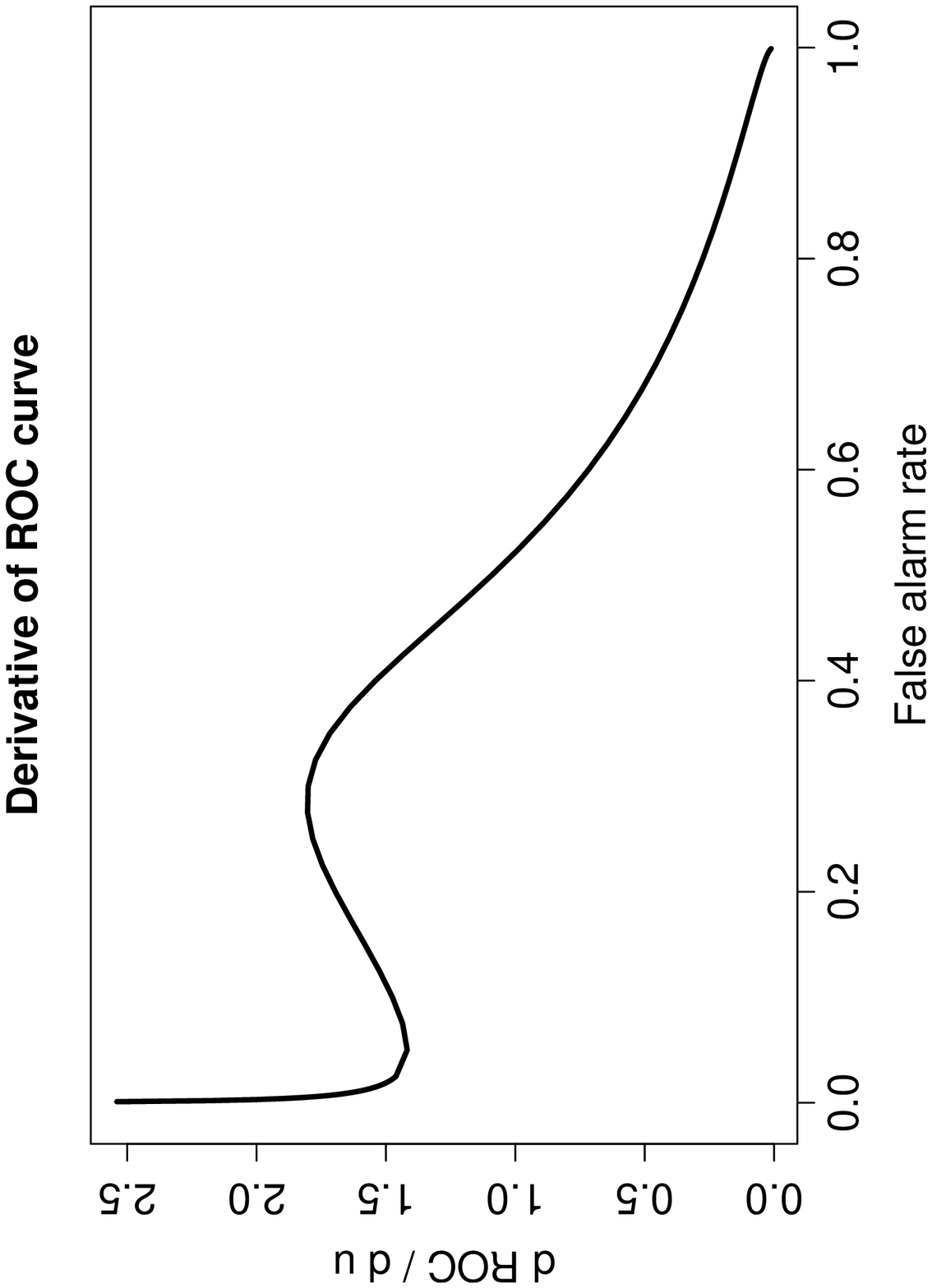}}
\end{turn}
\fi
%\end{center}
\end{figure}
\paragraph{Area under the curve (AUC).} As a further measure of discriminatory
power the \emph{Area under the curve} (AUC) is defined as the area between the
ROC curve and the axis of abscissa in Figure \ref{fig:4}. This area can be
calculated as the integral of the ROC curve from 0 to 1, i.e.
\begin{subequations}
\begin{equation}\label{eq:AUC_1}
  AUC\ = \ \int_0^1 ROC(u)\, d u.
\end{equation}
Alternatively, the AUC can be described as a probability,
    namely that the score of a non-defaulter selected at random is higher than
    the score of an independently selected defaulting borrower. Hence,
\begin{equation}\label{eq:Mann_Whitney}
  AUC\ = \ \mathrm{P}[S_D < S_N],
\end{equation}
where $S_N$ and $S_D$ are independent and distributed according to $F_N$ and
$F_D$ respectively. Moreover, it
    can be proved \citep[cf., for instance][]{Engelmannetal03}
     that the AUC is just an affine transformation of AR, namely
\begin{equation}\label{eq:AUC_AR}
    AUC \quad =\quad \frac{AR+1}{2}.
\end{equation}
As a consequence of this last observation, the higher AUC the higher
    is the discriminatory power of the rating system under consideration,
    as is the case for AR. Moreover, maximizing AUC is equivalent to maximizing discriminatory power as
maximizing the area under the ROC curve due \eqref{eq:ROC_deriv}
and \eqref{eq:Bayes} results in high PDs
for small scores and low PDs for large scores. Additionally, \eqref{eq:AUC_AR} shows
that the value of AR -- like the value of AUC -- depends on the conditional
densities of the score variable given the state of the borrower but not on the
total probability of default in the portfolio.
\end{subequations}

There is also an important consequence from the representation of AUC as a
    probability. The non-parametric Mann-Whitney test \citep[see, e.g.][]{Sheskin97}
    for the hypothesis that one distribution
    is stochastically greater than another can be applied as a test on whether there
    is discriminatory power at all or not. Additionally, a Mann-Whitney-like test
    for comparing the discriminatory power values of two or more rating systems
    is available \citep[cf.][]{Engelmannetal03}.

\paragraph{Error rates as measures of discriminatory power.} We have seen that
the ROC curve may be interpreted as a ``type I error level''-power diagram related
to cut-off decision rules in the sense of \eqref{eq:cut_I} and \eqref{eq:cut_II}, based
on the score variable under consideration. Another approach to measuring discriminative
power is to consider only total probabilities of error instead of type I and II error probabilities
separately.

The first example of an error-rate based measure of discriminatory power is the
\emph{Baysian error rate}. It is defined as the minimum total probability of error
that can be reached when cut-off rules are applied.
\begin{subequations}
  \begin{align}
    \text{Baysian error rate} & = \min_s \mathrm{P}[\text{Erroneous decision when
    cut-off rule with threshold $s$ is applied}]\notag\\
    &= \min_s \bigl(\mathrm{P}[Z=D]\,\mathrm{P}[S>s\,|\,Z=D] +
            \mathrm{P}[Z=N]\,\mathrm{P}[S\le s\,|\,Z=N]\bigr)\notag\\
    &= \min_s \left(p\,\bigl(1-F_D(s)\bigr) + (1-p)\,F_N(s)\right).\label{eq:Baysian}
  \end{align}
  In the special case of a hypothetical total PD of 50 percent
    the Baysian error rate is called \emph{classification error}. Assume
    that defaulters tend to receive smaller scores than non-defaulters, or,
    technically speaking, that $F_D$ is stochastically smaller than $F_N$ (i.e.\ $F_D(s) \ge F_N(s)$ for
    all $s$). The classification error can then
    be written as
  \begin{equation}\label{eq:classification_error}
    \text{Classification error}\ =\ 1/2 - 1/2\,\max_s |F_D(s) - F_N(s)|.
  \end{equation}
  The maximum term on the right-hand side of \eqref{eq:classification_error} is
  just the population version of the well-known \emph{Kolmogorov-Smirnov} statistic
  for testing whether the two distributions $F_D$ and $F_N$ are identical.
  The conditional distributions of the score variable being identical means that
  the score variable has not any discriminatory power. Thus, the classification
  error is another example of a measure of discriminatory power for which
  well-known and efficient test procedures are available.
  The so-called \emph{Pietra-index} reflects the maximum distance of a ROC curve and
  the diagonal. In the case where the likelihood ratio $f_D/f_N$ is a monotonous
  function,
  the Pietra-index can be written as an affine transformation of the Kolmogorov-Smirnov statistic
  and therefore is equivalent to it in a statistical sense.

  If the likelihood ratio is monotonous, the Kolmogorov-Smirnov statistic
  has an alternative representation as follows:
  \begin{equation}\label{eq:KS_alternative}
    \max_s |F_D(s) - F_N(s)|\,=\,1/2 \int_{-\infty}^\infty |f_D(s) - f_N(s)|\, d s \,\in\,[0,1/2].
  \end{equation}
    This representation is interesting because it allows to compare the Kolmogorov-Smirnov
    statistic with the \emph{information value}, a discrepancy measure which is based on relative entropies.
    We will not explain here in detail the meaning of relative entropy. What is important here, is the
    fact that the information value can be written in a way that
    suggests to interpret the information value as something like a ``weighted Kolmogorov-Smirnov''
    statistic.
  \begin{align}\label{eq:IV}
    \text{\hypertarget{IV}{Information value}} & = \mathrm{E}\bigl[
    \log\frac{f_D(S)}{f_N(S)}\,|\,D\bigr] + \mathrm{E}\bigl[
    \log\frac{f_N(S)}{f_D(S)}\,|\,N\bigr]\\
& = \int_{-\infty}^\infty (f_D(s) - f_N(s))(\log f_D(s) - \log f_N(s))\,d s\notag\\
    &\in\,[0,\infty).\notag
  \end{align}
Note that the information value is also called \emph{divergence} or \emph{stability index}.
    Under the notion stability index it is sometimes used as a tool to monitor the stability
    of score variables over time.
\end{subequations}

\paragraph{Measuring discriminatory power as variation of the PD conditional
on the score.}
So far we have considered measures of discriminatory power which are
    intended to express the discrepancy of the conditional distributions of
    the scores for the defaulters population and the non-defaulters population
    respectively. Another philosophy of measuring discriminatory power is based
    on measuring the variation of the conditional PD given the scores.
    Let us first
    consider the two extreme cases.

A score variable has no discriminatory power at all if the two conditional densities
of the score distribution (as illustrated in Figure \ref{fig:1}) are identical. In
that case the borrowers' score variable $S$ and state variable $Z$ are stochastically
independent. As a consequence, the conditional PD given the score is constant and equals
the total PD.
\begin{subequations}
  \begin{equation}\label{eq:PD_constant}
\mathrm{P}[D\,|\,S]\, =\, p.
  \end{equation}
One could also say that  the
score variable $S$ does not bear any information about potential default.
Obviously, such a score variable would be considered worthless.

The other extreme case is the case where the conditional PD given the
    scores takes on the values 0 and 1 only.
  \begin{equation}\label{eq:PD_full}
\mathrm{P}[D\,|\,S]\, =\, \mathbf{1}_D \, =\, \left\{%
\begin{array}{ll}
    1, & \hbox{if borrower defaults;} \\
    0, & \hbox{if borrower remains solvent.} \\
\end{array}%
\right.
  \end{equation}
    This would be an indication of
    a perfect score variable as in such a case there were no uncertainty about
    the borrowers' future state any more. In practice, none of these two extreme cases will occur. The conditional
    PD given the score will in general neither take on the values 0 and 1 nor will it
    be constant either.
\end{subequations}

In regression analysis, the determination coefficient $R^2$ measures
    the extent to which a set of explanatory variables can explain the variance
    of the variable which is to be predicted. A score variable or the grades of
    a rating system may be considered explanatory variables for the default state
    indicator. The conditional PD given the score is then the best predictor of
    the default indicator by the score in the sense of \eqref{eq:least_squares}. 
    Its variance can be compared to the
    variance of the default indicator in order to obtain an $R^2$ for this special
    situation.
    \begin{subequations}
  \begin{equation}\label{eq:R2}
    R^2 = \frac{\mathrm{var}[\mathrm{P}[D\,|\,S]]}{\mathrm{var}[\mathbf{1}_D]}
    = \frac{\mathrm{var}[\mathrm{P}[D\,|\,S]]}{p\,(1-p)} = 1 -
    \frac{\mathrm{E}\bigl[(\mathbf{1}_D -
    \mathrm{P}[D\,|\,S])^2\bigr]}{p\,(1-p)} \in [0,1].
  \end{equation}
The closer the value of $R^2$ is to one, the better the score $S$ can explain
    the variation of the default indicator. In other words, if $R^2$ is close to one,
    a high difference in the score values does more likely indicate a corresponding
    difference in the values of the default indicator variable.
    Obviously, maximizing $R^2$ is equivalent to maximizing $\mathrm{var}[\mathrm{P}[D\,|\,S]]$ and to
    minimizing $\mathrm{E}\bigl[(\mathbf{1}_D -
    \mathrm{P}[D\,|\,S])^2\bigr]$.

    The sum over all borrowers of the squared differences of the default indicators and the conditional
    PDs  given the scores  divided by the sample size is called \emph{Brier score.}
    \begin{equation}\label{eq:Brier}
    \text{Brier score}\ =\ \frac 1 n \sum_{i=1}^n\bigl(\mathbf{1}_{D_i} -
  \mathrm{P}[D\,|\,S= S_i]\bigr)^2.
\end{equation}
The Brier
score is a natural estimator of $\mathrm{E}\bigl[(\mathbf{1}_D -
    \mathrm{P}[D\,|\,S])^2\bigr]$ which is needed for calculating the
    $R^2$ of the score variable under consideration. Note that as long as
    default or non-default of borrowers cannot be predicted with certainty
    (i.e.\ as long as \eqref{eq:PD_full} is not satisfied) $\mathrm{E}\bigl[(\mathbf{1}_D -
    \mathrm{P}[D\,|\,S])^2\bigr]$ will not equal 0.

    In practice, the development of a rating system or score variable involves both an
    optimization procedure (such as maximizing $R^2$) and an estimation exercise (estimating
    the PDs given the scores $\mathrm{P}[D\,|\,S=s]$). The Brier score can be used for both
    purposes. On the one hand selecting an optimal score variable may be
    conducted by minimizing $\mathrm{E}\bigl[(\mathbf{1}_D -
    \mathrm{P}[D\,|\,S])^2\bigr]$, which usually also involves estimating $\mathrm{P}[D\,|\,S=s]$ for all
    realizable score values. On the other hand, when the score variable $S$ has already
    been selected, the Brier
    score may be used for calibration purposes (see Section \ref{se:calibration}).
\end{subequations}
\refstepcounter{figure}
\begin{figure}[ht]
%  \begin{center}
\centering
  \parbox{12.0cm}{Figure \thefigure:
  \emph{Graph of function $p \mapsto - \bigl(p\,\log p +
  (1-p)\,\log (1-p)\bigr)$.}}
\label{fig:6}%\\[2ex]
\ifpdf
    \resizebox{\height}{8.0cm}{\includegraphics[width=12.8cm]{entropy.pdf}}
\else
\begin{turn}{270}
\resizebox{\height}{12.8cm}{\includegraphics[width=8.0cm]{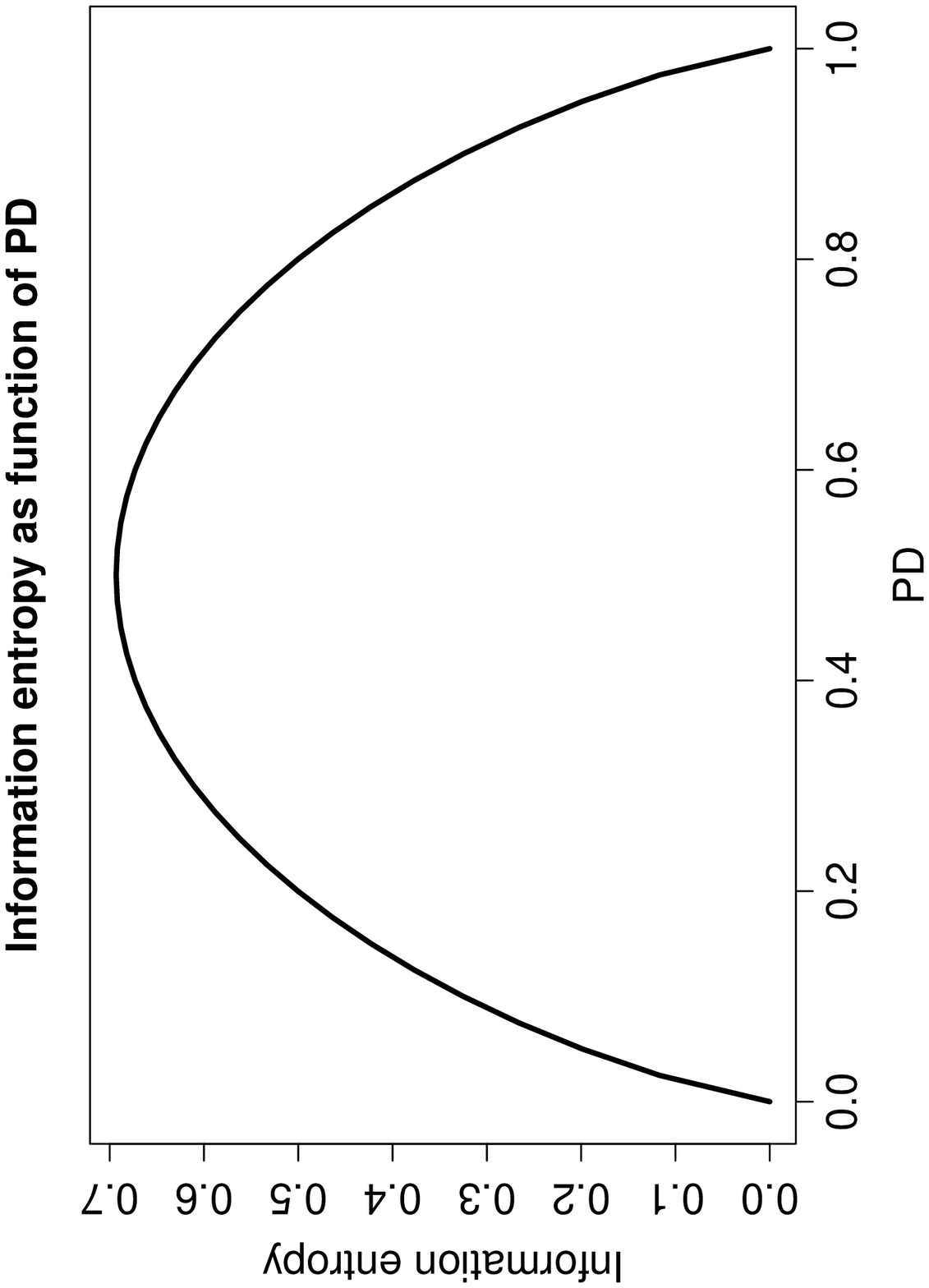}}
\end{turn}
\fi
%\end{center}
\end{figure}
\paragraph{Further entropy measures of discriminatory power.} Besides the information value
defined in \eqref{eq:IV} sometimes also other entropy-based measures of discriminatory
power are used in practice.

\begin{subequations}
For any event with probability $p$ its \emph{information entropy} is
    defined as
    \begin{equation}\label{eq:IE}
    H(p)\ =\ - \bigl(p\,\log p +
  (1-p)\,\log (1-p)\bigr)
\end{equation}
Note from Figure \ref{fig:6} that
$H(p)$ is close to 0 if and only if $p$ is close to 0 or close to 1.
As a consequence, information
    entropy can be regarded as a measure of uncertainty of the underlying event.
When discriminatory power of a score variable has to be measured,
    it can be useful to consider the information entropy applied to the
    conditional PD given the scores, i.e.\ $H\bigl(\mathrm{P}[D\,|\,S]\bigr)$.
    If then the average value of the information
    entropy is close to zero, the conditional PD given the scores will be close
    to zero or to one in average, indicating high discriminatory power.
    Formally, the average information entropy of the conditional PD is
    described as \emph{conditional entropy} $H_S$ which is defined as the expectation
    of the information entropy applied to the conditional PD given the scores.
\begin{equation}\label{eq:CE}
    H_S\ =\ \mathrm{E}\bigl[H\bigl(\mathrm{P}[D\,|\,S]\bigr)\bigr].
\end{equation}
As both the conditional PD given the scores as well as the calculation
    of the expectation depend on the portfolio-wide total PD, it is not sensible
    to compare directly the conditional entropy values of score variables from
    populations with different portions of defaulters. However, it can be shown
    by Jensen's inequality that the conditional entropy never exceeds the information
    entropy of the total probability of default of the population under consideration.
    Therefore, by using the \emph{conditional information entropy ratio} (CIER),
    defined as ratio of information entropy of the total PD minus conditional entropy
    of the conditional PDs and the information entropy of the total PD,
    conditional entropy values of different score variables can be made commensurable.
\begin{equation}
    \text{CIER}\ =\ \frac{H(p) - H_S}{H(p)} \,\in\,[0,1]
  \end{equation}
The closer the value of CIER is to one, the more information about default
    the score variable $S$ bears, in the sense of providing conditional PDs given the
    scores which are close to 0 or to 1.
\end{subequations}

%%%%%%%%%%%%%%%
% New section %
%%%%%%%%%%%%%%%
\section{Calibration of rating systems}
\label{se:calibration}

The issue with calibration of rating systems or score variables
    is how accurate the estimates of the conditional default probability
  given the score are. Supervisors, in particular, require that the
  estimates are not too low when they are used for determining
  regulatory capital requirements. In the following, we will consider
  some tests on calibration that are conditional
  on the state of the economy. These are the binomial test, the Hosmer-Lemeshow
  test and the Spiegelhalter test. As an example for unconditional tests,
  we will then discuss  a normal approximate test.

\paragraph{Conditional versus unconditional tests.} The notions of conditional
and unconditional tests in the context of validation for Basel II can be
best introduced by relating these notions to
    the notions of PIT and TTC PD estimates.

PD estimates can be based (or, technically speaking, conditioned) on the current
    state of the economy, for instance
by inclusion of macro-economic co-variates in a regression process. The co-variates
are typically the growth rate of the gross domestic product, the unemployment rate
or similar indices. The resulting PD estimates are then
called \emph{Point-In-Time} (PIT). With such estimates,
given an actual realization of the co-variates, an assumption of independence
of credit events may be adequate, because most of their dependence might have been captured
by incorporating the economic state variables in the PDs estimates.

In contrast, unconditional PD estimates are not based on a current state of the economy.
Unconditional PDs that are estimated based on data from a complete economic
cycle are called \emph{Through-The-Cycle} (TTC). When using unconditional PDs,
no assumption of independence can be made, since
then the variation of the observed default rates cannot be any longer explained by
the variation of conditional PDs which are themselves random variables.

\paragraph{Binomial test.} Consider one fixed rating grade specified by a range
    $s_0 \le S \le s_1$, as described, for instance, in
    \eqref{eq:limit1} and \eqref{eq:mapping_1}.
    It is then reasonable to assume that an average PD $q$ has been forecast for the rating
    grade under consideration. Let $n$ be the number of borrowers
    that have been assigned this grade.

    If the score variable is able to reflect to some extent the current state
    of the economy, default events among the borrowers may be considered stochastically
    independent. Under such an independence assumption, the number of defaults in the rating grade is
  binomially distributed with parameters $n$ and $q$.
  Hence the \emph{binomial test} \citep[cf., e.g.][]{Brown01}
  may be applied to test the hypothesis ``the true PD of this grade
    is not greater than the forecast $q$''. If the number of borrowers within the grade and the
    hypothetical PD $q$ are not too small, thanks to the central limit theorem under
    the hypothesis the binomial distribution  can be approximated with a normal distribution.
    As already mentioned, for this approximation to make sense is important that the independence assumption
    is justified. This will certainly not be the case when the PDs are estimated
    through-the-cycle. The following example illustrates what then may happen.

    \paragraph{Example.} Assume that 1000 borrowers have been assigned 
    the rating grade under consideration. The bank forecasts for this grade a PD of 1 percent.
    One year after the forecast 19 defaults are observed.

If we assume independence of the default events,
    with a PD of 1 percent the probability to observe 19 or more defaults
    is 0.7 percent. Hence, the hypothesis that the true PD is not
    greater than 1 percent can be rejected with 99 percent confidence.
    As a consequence, we would conclude that the bank's forecast was too
    optimistic.

    Assume now that the default events are not independent. For the purpose of illustration,
    the dependence then can be modelled by means of a normal copula with uniform correlation 5 percent
    \citep[see, e.g.][for details of the one-factor model]{PT05}.
    Then, with a PD of 1 percent, the probability to observe 19 or more defaults is 11.1 percent.
    Thus, the hypothesis that the true PD is not
    greater than 1 percent cannot be rejected with 99 percent confidence.
    As a consequence, we would accept the bank's forecast as adequate.

\paragraph{Hosmer-Lemeshow test.} The binomial test
    can be appropriate to check a single PD forecast. However, if --
    say -- twenty PDs of rating grades are tested stand-alone, it is quite
    likely that at least one of the forecasts will be erroneously rejected.
    In order to have at least control over the probability of such
    erroneous rejections, joint tests for several grades have to be used.

    So, assume that there are PD forecasts $q_1, \ldots, q_k$ for rating
grades $1, \ldots, k$.
Let $n_i$ denote the number of borrowers with grade $i$ and
$d_i$ denote the number of defaulted borrowers with grade $i$.
The \emph{Hosmer-Lemeshow statistic} $H$ for such a sample is the
    sum of the squared differences of forecast and observed numbers
    of default, weighted by the inverses of the theoretical variances
    of the default numbers.
  \begin{equation}
    H\ =\ \sum_{i=1}^k \frac{(n_i\,q_i-d_i)^2}{n_i\,q_i\,(1-q_i)}.
  \end{equation}
Under the usual assumptions on the appropriateness of normal
    approximation (like independence, enough large sample size), the Hosmer-Lemeshow
    statistic is $\chi_k^2$-distributed
    under
    the hypothesis that all the PD forecasts match the true PDs. This fact
    can be used to determine the critical values for testing the hypothesis
    of having matched the true PDs. However, also for the Hosmer-Lemeshow test, the assumption of
    independence is crucial. Additionally, there may be an issue of
    bad approximation for rating grades with small numbers of borrowers.

\paragraph{Spiegelhalter test.} If the PDs of the borrowers are individually estimated, both the
    binomial test and the Hosmer-Lemeshow test require averaging the
    PDs of borrowers that have been assigned the same rating grade.
    This procedure can entail some bias in the calculation of
    the theoretical variance of the number of defaults.
    With the Spiegelhalter test, one avoids this problem.
    As for the binomial and Hosmer-Lemeshow test, also for the Spiegelhalter
    test independence of the default events is assumed.
    As mentioned earlier, if the PD are estimated point in time,
    the independence assumption may be justified.

\begin{subequations}
We consider borrowers $1, \ldots, n$ with scores $s_i$ and PD estimates $p_i$.
Given the scores, the borrowers are considered to default or remain solvent
independently. Recall the notion of Brier score from \eqref{eq:Brier}. In contrast
to the situation when a rating system or score variable is developed, for
the purpose of validation we assume that realizations of the ratings are given
and hence non-random. Therefore we can drop the conditioning on the score
realizations in the notation. In the context of validation, the Brier score
is also called
\emph{Mean squared error (MSE)}.
\begin{equation}
    MSE \ =\ 1/n \sum_{i=1}^n (\mathbf{1}_{D_i}-p_i)^2,
\end{equation}
where $\mathbf{1}_{D_i}$ denotes the default
indicator as in \eqref{eq:PD_full}. The null hypothesis for the test is
``all PD forecasts match exactly the true
conditional PDs given the scores'', i.e.\ $p_i = \mathrm{P}[D_i\,|\,S_i=s_i]$ for all $i$.

It can be shown that under the null we have
\begin{align}
    \mathrm{E}[MSE] & = 1/n \sum_{i=1}^n p_i\,(1-p_i)\qquad\text{and}\\
    \mathrm{var}[MSE] & = n^{-2} \sum_{i=1}^n p_i\,(1-p_i)\,(1-2\,p_i)^2.
\end{align}
Under the assumption of independence given the score values,
    according to the central limit theorem, the distribution of the
    standardized mean squared error
\begin{equation}
    Z \ = \ \frac{MSE - \mathrm{E}[MSE]}{\sqrt{\mathrm{var}[MSE]}}
\end{equation}
is approximately standard normally distributed under the null.
Thus, a joint test of the hypothesis ``the calibration of the PDs with respect
to the score variable is correct'' can be conducted \citep[see][for examples
from practice]{RS05}.
\end{subequations}

\paragraph{Testing unconditional PDs.} As seen before
    by example,
    for unconditional PD estimates assuming independence
    of the defaults for testing the adequacy of the
    estimates could result in too conservative tests.
    However, if a time-series of default rates is available,
    assuming independence over time might be justifiable.
    Taking into account that unconditional PD estimates usually are constant\footnote{%
    \citet{Bloch04} provide a modification of the test for the case of non-constant PD estimates.
    } over time,
    a simple test can be constructed that does not involve any assumption of
    cross-sectional independence among the borrowers within a
    year.
We consider a fixed rating grade with $n_t$ borrowers (thereof $d_t$ defaulters) in
  year $t=1, \ldots, T$. Additionally, we assume that the estimate $q$ of the PD
    common to the borrowers in the grade is of TTC type and constant over time,
  and that defaults in different years are independent.
In particular, then the annual default rates $d_t/n_t$ are realization of independent
random variables.
The standard deviation $\sigma$ of the default rates can in this case be estimated
with the usual unbiased estimator
\begin{subequations}
  \begin{equation}
    \hat{\sigma}^2 \ =\ \frac 1{T-1} \sum_{t=1}^T \left(\frac{d_t}{n_t} - \frac
    1 T \sum_{\tau=1}^T \frac{d_\tau}{n_\tau}\right)^2.
  \end{equation}
If the number  $T$ of observations is not too small, and
    under the hypothesis that the true PD is not greater than $q$,
    the standardized average default rate is approximately standard
    normally distributed. As a consequence, the hypothesis should be
    rejected if the average default rate is greater than $q$ plus
    a critical value derived by this approximation. Formally,
    reject ``true PD $\le\ q$'' at
  level $\alpha$ if
  \begin{equation}\label{eq:normal_test}
    \frac 1 T \sum_{\tau=1}^T \frac{d_\tau}{n_\tau} \ >\ q +
    \frac{\hat{\sigma}}{\sqrt{T}}\,\Phi^{-1}(1-\alpha).
  \end{equation}
  \end{subequations}
  As mentioned before, the main advantage of the \emph{normal test} proposed here
    is that no assumption on cross-sectional independence is needed.
    Moreover, the test procedure seems even to be robust against
    violations of the assumption of inter-temporal independence, in the
    sense that the test results still appear reasonable when there is
    weak dependence over time. More critical appears the assumption
    that the number $T$ of observations is large. In practice,
    time series of length five to ten years do not seem to be uncommon.
    In  Tables 1 and 2 we present the results of an
    illustrative Monte-Carlo simulation exercise in order to give an impression
of the impact of having a rather short time series.

\begin{minipage}{7cm}
\begin{center}
\emph{Table 1}\\
\emph{Estimated PD = 2\%, $T=5$, $\alpha=1\%$}\\[1ex]
\begin{tabular}{|@{\ }c@{\ }|@{\ }c@{\ }|}
\hline
True PD & Rejection rate\\ \hline\hline
1.0\% & 0.00\%\\ \hline
1.5\% & 0.01\%\\ \hline
2.0\% & 2.05\%\\ \hline
2.5\% & 19.6\%\\ \hline
5.0\% & 99.2\% \\ \hline
\end{tabular}
\end{center}
\end{minipage}\hfill
    \begin{minipage}{7cm}
\begin{center}
\emph{Table 2}\\
\emph{Estimated PD = 2\%, $T=5$, $\alpha=10\%$}\\[1ex]
\begin{tabular}{|@{\ }c@{\ }|@{\ }c@{\ }|}
\hline
True PD & Rejection rate\\ \hline\hline
1.0\% & 0.00\%\\ \hline
1.5\% & 0.60\%\\ \hline
2.0\% & 7.96\%\\ \hline
2.5\% & 30.1\%\\ \hline
5.0\% & 99.2\%\\ \hline
\end{tabular}
\end{center}
    \end{minipage}

The exercise whose results are reflected in Tables 1 and 2
was conducted in order to check the quality of the normal approximation
for the test of the unconditional PDs according to \eqref{eq:normal_test}.
For two different type I error probabilities the tables present the true rejection
rates of the hypothesis ``true PD not greater than 2 percent'' for different values
of the true PDs. By construction of the test, the rejection rates ought to be
not greater than the given error probabilities as long as the true PDs are not
greater than 2 percent. For the smaller error probability of 1 percent this seems
to be a problem, but not a serious one. However, the tables also reveal that
the power of the test is rather moderate. Even if the true PD is so clearly
greater than the forecast PD as in the case of 2.5 percent, the rejection
rates are 19.6 and 30.1 percent respectively only.

%%%%%%%%%%%%%%%
% New section %
%%%%%%%%%%%%%%%
\section{Conclusions}
\label{se:conclusions}

With regard to measuring discriminatory power, the Accuracy Ratio and the Area under
    the Curve seem promising\footnote{%
The selection of the topics and the point of view taken in this chapter is primarily
    a regulatory one. This is caused by the author's background in a regulatory
    authority. However, the presentation does not reflect any official
    regulatory thinking. The regulatory bias should be kept in mind  when the following conclusions
    are read. A procedure which may be valuable for regulatory
    purposes need not necessarily also be appropriate for bank-internal applications.
    } tools since their statistical properties are well investigated
    and they are available together with many auxiliary features in most of the more
    popular statistical software packages.

    With regard to testing calibration, for conditional PD estimates powerful
    tests like the binomial, the Hosmer-Lemeshow and the Spiegelhalter test are
    available. However, their appropriateness strongly depends on an independence assumption
    which needs to be justified on a case by case basis. Such independence assumptions
    can at least partly be avoided, but at the price of losing power as illustrated with
    a test procedure based on a normal approximation.

%%%%%%%%%%%%%%
% References %
%%%%%%%%%%%%%%

%
\end{document}